\documentclass[jcp,twocolumn,groupedaddress
reprint,
superscriptaddress,
showpacs,preprintnumbers,
amsmath,amssymb,
aps,
]{revtex4}
\usepackage{graphicx}
\usepackage{dcolumn}
\usepackage{bm}%
\usepackage[colorlinks=true,citecolor=blue]{hyperref}
\hypersetup{colorlinks=true,citecolor=blue,linkcolor=red,urlcolor=blue}

\usepackage{color}
\definecolor{Green}{RGB}{0,204,102}
\definecolor{Purple}{RGB}{102,0,255}
\definecolor{Blue}{RGB}{51,153,255}
\definecolor{Red}{RGB}{151,010,010}

	\newcommand {\be}{\begin{equation}}
	\newcommand {\ee}{\end{equation}}
	\newcommand {\bea}{\begin{array}}
	
	\newcommand {\eea}{\end{array}}

	\newcommand {\ep}{\epsilon}

	\newcommand{\tqr}{\textquotedblright}
	\newcommand{\tql}{\textquotedblleft}		

\begin{document}

\title{Phosphorene as a nanoelectromechanical material}

\author{Zahra Nourbakhsh}
\affiliation{School of Nano Science, Institute for Research in Fundamental Sciences (IPM), Tehran 19395-5531, Iran}
\author{Reza Asgari}
\affiliation{School of Nano Science, Institute for Research in Fundamental Sciences (IPM), Tehran 19395-5531, Iran}
\affiliation{School of Physics, Institute for Research in Fundamental Sciences (IPM), Tehran 19395-5531, Iran}

\date{\today}					

\begin{abstract}
Based on density functional simulations combined with the Landauer transport theory, the mechanical strain impacts on the chemical bonds of phosphorene and their effects on the electronic properties are studied. Moreover, the effect of the tensile strain along the zigzag direction on the charge transport properties of a two-terminal phosphorene device is evaluated.
Enhancement of the intra-planar interactions, in particular between the next-nearest-neighbors in strained phosphorene is found to be essential in the band structure evolution.
The charge transport analyzing shows that phosphorene has a strong piezoconductance sensitivity, which makes this material highly desirable for high-pressure nanoelectromechanical applications.
The piezoconductance gauge factor increases by strain from 46 in 5\% tension to 220 in 12\% tension which is comparable to state-of-the-art silicon strain sensors. The transmission pathways monitor the current flowing in terms of the chemical bonds and hopping, however, the transport mostly arises from the charge transferring through the chemical bonds.
The strong anisotropy in the transport properties along zigzag and armchair directions is observed.
\end{abstract}

\maketitle

\section{Introduction}\label{sec:intro}

Strain is an old but alive concept in the semiconductor technology \cite{StrBook}. Strain engineering is a well-known technique to boost the device performance. Electrical, optical, magnetic, thermal, and chemical properties could be tuned and controlled by external forces \cite{StrBook}. Based on the higher elasticity and larger holding strain limits of nanomaterials compared to their bulk structure \cite{nanoSt1,nanoSt2}, they attract much attention for strain engineering applications. Furthermore, low dimensional materials are important in application to reduce the size of the electronic devices in a modern technology.

Among two-dimensional (2D) crystalline materials, phosphorene, which is an elemental 2D material, shows great potential for nanoelectronic applications.
Phosphorene is a promising 2D material to utilize as an ideal electronic device. It has an $1.4$~eV direct band gap, high carrier mobility and high on/off ratio for field-effect transistor applications \cite{ph-birth, aniso1, ph-app1, ph-app2, ph-app3}. Owing to $sp^3$ hybridization between phosphorus atoms in phosphorene, it is a semiconductor with a puckered honeycomb structure and remarkable in-plane anisotropic behavior between the armchair ($ac$) and zigzag ($zz$) directions \cite{aniso1, aniso2, PExcExp,nour1}.
Due to its puckered structure, phosphorene shows superior flexibility and mechanical properties compared to other 2D materials in both the $ac$ and $zz$ directions. In addition, phosphorene can withstand tensile strain up to 30\% \cite{strp1} which is higher than reported values for graphene ($\sim$ 15-20\% ~\cite{gr}) and MoS$_2$ ($\sim$ 11\% ~\cite{mos2}) which are suffering from plasticity and low fracture limit.

The great mechanical properties of phosphorene make strong motivation to study this system under various types of strain and deformations \cite{blackp2,asgari,strp2,strp3,strp4}. The Infrared spectroscopy as well as theoretical investigations reveal the significant sensitivity and tuning the gap size of phosphorene and its carriers effective masses by imposing external strains.
In addition, strain can convert the direct band gap to indirect one, and also induces a semiconductor-metal transition in phosphorene \cite{strp2}. The stability of the strained phosphorene system has been investigated using the $ab~initio$ phonon dispersion calculations \cite{negPhonon}.
The results indicate that phosphorene structure is stable under the tensile strains, but it shows dynamical instability and softening against an even small amount of the compression.
Also, experimental and theoretical studies reveal that under an appropriate in-plane uniaxial or biaxial tensile strain, the preferred conducting direction is rotated from the $ac$ to $zz$ \cite{aniso2}.

Based on the particular mechanical properties of phosphorene and the observed coupling between its mechanical and electronic properties, phosphorene could be considered as a favorable novel material for nanoelectromechanical systems applications. Our aim in this article is to investigate the piezoresistance/piezoconductance characteristic of phosphorene under a uniaxial tension strain along the $zz$ direction. Piezoresistance, which means changing the electrical resistance of a system as a result of pressure, was discovered by lord Kelvin in elongated copper and iron systems in 1856 \cite{piezoRrev}. High piezoresistivity has been explored experimentally in silicon and germanium bulk structures in 1954 \cite{smith}. These materials are widely used as commercial piezoresistance devices \cite{piezoRrev}. The magnitudes and signs of the
piezoresistive coefficients strongly depend on the electron mobility, crystallographic direction, stress and to the crystallographic axes~\cite{piezoRrev}. Piezoresistance has been shown experimentally in graphene \cite{gr} and MoS$_2$ \cite{mos2-piezo} monolayers. The piezoresistance coefficient in MoS$_2$ is two orders of magnitude higher than that of reported for graphene, and it is in the same order of the sensitivity of the best Si and Ge piezoresistance devices. Also, piezoresistivity has been measured for the bulk black phosphorous, first in 1964 \cite{blackp1} and very recently by Zhang $et~al.$ \cite{blackp2}.

In this study, we consider a phosphorene device under an applied external strain and calculate its electronic structure and its $I-V$ curve based on the $ab~initio$ Kohn-Sham density functional theory (DFT) \cite{k-sh} together with nonequilibrium Green's function (NEGF) \cite{negf} methods stemming from the successful Landauer formalism \cite{landauer}.
This method provides a good agreement between theoretical results and those measured in experiments \cite{exp1} and has been widely applied to evaluate the transport properties of different mesoscopic systems.

Our calculations predict a strong piezoconductance sensitivity in our proposed phosphorene based device. This property could be important for a wide range of  applications in biological, environmental, medical, chemical pressure sensors, and transducer devices \cite{piezoRrev}. Furthermore, the modern piezoresistance/piezoconductance materials could satisfy the technological desire for smaller size and higher performance electronic devices.

In addition, using Bader's \tql atoms in molecules\tqr  theory \cite{bader}, we present a novel description of the effect of bonding in the band structure of strained phosphorene systems. Besides, on the basis of $ab~initio$ calculations, we explain the reason for the changing of the preferred conducting direction in the strained systems.

This paper is organized as follows. Having reviewed the theoretical and technical methods in the next section, we present our results in Sec.~\ref{sec:results}. Analyzing and description of the band structure of phosphorene and doped phosphorene under strain, calculating transport properties of different strained phosphorene-based systems under an applied bias, the $I-V$ characteristic, the transmission spectrum, the carrier type, and the transmission pathways are discussed in this section.
Finally, we end with a brief summary of our main results in Sec.~\ref{sec:summary}.

\section{Methodology}\label{sec:method}

\begin{figure}
\centering
\includegraphics[width=0.9\linewidth]{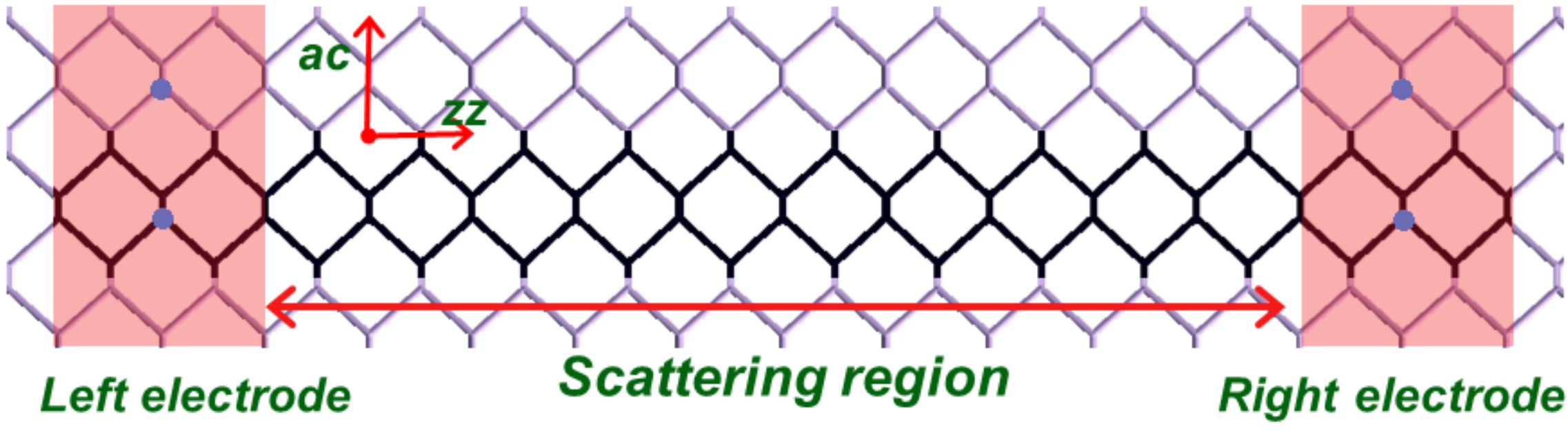}
\caption{\label{device} (Color online) The darker area shows the computational setup for a two-terminal phosphorene device.  The transport direction is along the $zz$ direction and the periodic boundary conditions where represent by the lighter color, are available along the $ac$ direction. The electrodes have a semi-infinite nature and they repeat in the left and right sides. The system includes three parts: semi-infinite left and right electrodes, and the central scattering region. The balls in the electrodes denote the dopant impurity atoms. The shaded areas show the left and right electrode supercells containing two primitive unit cells with the length of $\sim 0.7$~nm. This length should be large enough which electrode orbitals only interact with the nearest-neighbor cell in the repeating system.}
\end{figure}

We consider a phosphorene based system that consists of three main parts involving semi-infinite left and right electrodes, and a finite central scattering region.
Figure~\ref{device} shows the schematic representation of a two-terminal phosphorene device.
Under a bias voltage, the electrodes are at two different electrochemical potentials. Therefore, the charge carriers from the source electrode are partially transmitted through the scattering region and lastly absorbed in the drain electrode. Moreover, the reflected carriers come back to the source.
The main difficulty in the $ab~initio$ modeling of such a system is the non-periodicity of the device. Due to the existence of two electrodes and an externally applied bias, scattering boundary conditions have to be considered at the boundary between the electrodes and central scattering region. This issue can be resolved using the combination of non-equilibrium Green's function techniques with DFT. This method has been implemented in TranSIESTA software \cite{transiesta}, which uses a self-consistent solution of the Poisson equation and it is able to calculate the electron density of an open boundary system under an applied bias.

The key quantity to evaluate the transport properties is the transmission coefficient \cite{handbook} which is a function of the energy of the incoming carriers from the source electrode and bias voltage, $T(E,V_{Bias})$, and it is calculated using the retarded Green's function as $T(E,V) =  \Gamma_L(E,V) G(E,V) \Gamma_R(E,V) G^{\dag}(E,V)$, where $\Gamma_{L(R)}$ is the level broadening determined by $\Gamma_{L(R)}=i(\Sigma_{L(R)}-\Sigma^{\dag}_{L(R)})$, and $\Sigma_{L(R)}$ is the self-energy which defines the coupling between the semi-infinite left (right) electrode and the finite central region. The retarded Green's function of the central region is calculated as $G=[E+i\eta - H - \Sigma_L - \Sigma_R]^{-1}$, where $H$ is the Hamiltonian matrix of the central region. Regarding the standard
Landauer formalism, current is evaluated by transmission integration over the bias energy window. The relation at bias voltage $V$ is given by
\be
\label{landaue}
I(V)=\frac{G_0}{e}\int_{\mu_L}^{\mu_R}T(E,V)dE
\ee
where the coefficient $G_0=2e^2/h$ ($h$ and $e$ are the Planck constant and electron charge, respectively.) is known as the quantum of the conductance, and $\mu_{L(R)}=E_{\rm F}\pm eV/2$ is the electrochemical potentials of the left (right) electrode. The difference of the electrode potentials is equal to the value of the bias voltage, $eV=\mu_L-\mu_R$.

The $ab~initio$ DFT calculations are performed using SIESTA/TranSIESTA code \cite{siesta, transiesta}. To simulate the external mechanical strain, the lattice constant in the direction of the applied mechanical pressure is constrained, and in other directions is fully relaxed. The atomic positions are also relaxed in order that all components of all forces become less than $0.01$~eV/\AA. Nonlocal Troullier-Martins norm-conserving pseudopotentials are used to express the core electrons and the valence electrons are represented using the linear combinations of the atomic orbitals. We use a double-$\zeta$ polarized basis set within the generalized gradient approximation (GGA) in the scheme of Perdew, Burke, and Ernzerhof (PBE) \cite{pbe}. The real space Fourier expansion of the electron density is cut at 450~Ry. The Brillouin zone integrations are performed on the Monkhorst-Pack \cite{m-p} $k$-point grid of $16 \times 1 \times 14$, for the electronic structure calculations. It increases to $16 \times 1 \times 70$ during the transport calculations; such a large number of $k$ points is necessary to reproduce the semi-infinite electrodes. To simulate an isolated system, a vacuum gap of $20$~\AA~ is considered. The electron temperature is set at 300~K in the transport calculations. Our studied structures do not show any considerable magnetic moment. Since our calculations are in the low biases, we ignore the geometry optimization caused by the bias voltage.

In addition, we use AIMALL package \cite{aimall} to describe the bond properties of the system on the basis of Bader's \tql Atoms in Molecules\tqr  theory. In this scheme, the bond points are defined as the saddle points of the electronic charge density between two bound atoms. At a bond saddle point, the electron density displays a minimum in the bond direction and two maximum in the perpendicular directions. Furthermore, we employ several post-processing codes such as TBtrans (tight-binding transport)\cite{tbtran} and the sisl utility \cite{sisl}, as well as some home-made programming to extract different transport features of the system.

The lattice parameters of 2D phosphorene are obtained $a_{ac}=4.37$~\AA~ and $a_{zz}=3.25$~\AA. Furthermore, its band gap is $0.96$~eV which is close to one obtained using other DFT-PBE simulations\cite{strp1}.

\section{results and discussion}\label{sec:results}

\subsection{Bands and bonds in strained phosphorene}\label{sec:bond}

\begin{figure*}
	\centering
	\includegraphics[width=1\linewidth] {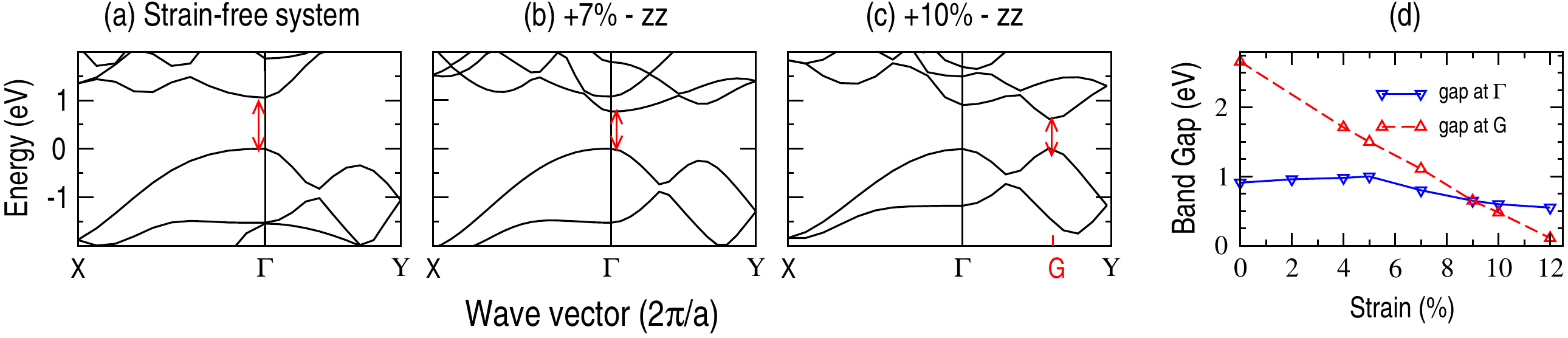}
	\caption{\label{band} (Color online) The electronic band structure of phosphorene in free conditions (a), and under two different tensile strains applied along the $zz$ direction  (b) and (c). The evolution of the band gap as a function of tension is plotted in the last panel.}
\end{figure*}

Figure~\ref{band} shows how the tensile strain along the $zz$ direction modulates the band structure of phosphorene. The gap size, as well as the gap position, are tuned by the tension. Moreover, strain changes the curvature of the band structure and consequently, it changes the carrier density and carrier effective masses especially for the conduction bands. The results presented in Fig.~\ref{band} are consistent with the experiment \cite{blackp2} and other simulated literature findings \cite{strp1}.
As the figure indicates, the band gap at the $\Gamma$ point slightly increases with increasing the tensile strain up to 5\%, afterwards it decreases with the tension enhancement. However, by further increasing of the strain, around the critical tensile strain of $\sim 9\%$, the gap position moves from the $\Gamma$ point to the $G$ point, located at $k_G \sim 0.6 \Gamma Y$ along the $ac$ direction meaning that the band gap is highly strain-tunable.
Shifting the gap position under strain is the feature of multi-valley semiconductors like silicon, germanium, and GaAs where strain could shift some valleys against others \cite{StrBook}. In what follows, we explain the underlying physics of this nonuniform band gap values in phosphorene in the presence of the tensile strain. The band gap at both $\Gamma$ and $G$ points remains direct and it decreases monotonously at the $G$ point with increasing tension, however, the semiconductor-metal transition has not been reported up to 14\% tension \cite{strp1}.
Another point in Fig.~\ref{band} is turning in the preferred conducting direction from $ac$ to $zz$ for the tensions above $\sim 7\%$. Accordingly, in the strained system, the conduction band in the $ac$ direction is no longer linear. The origin of this rotation will be discussed in the following.

\begin{figure}
	\centering
	\includegraphics[width=1\linewidth] {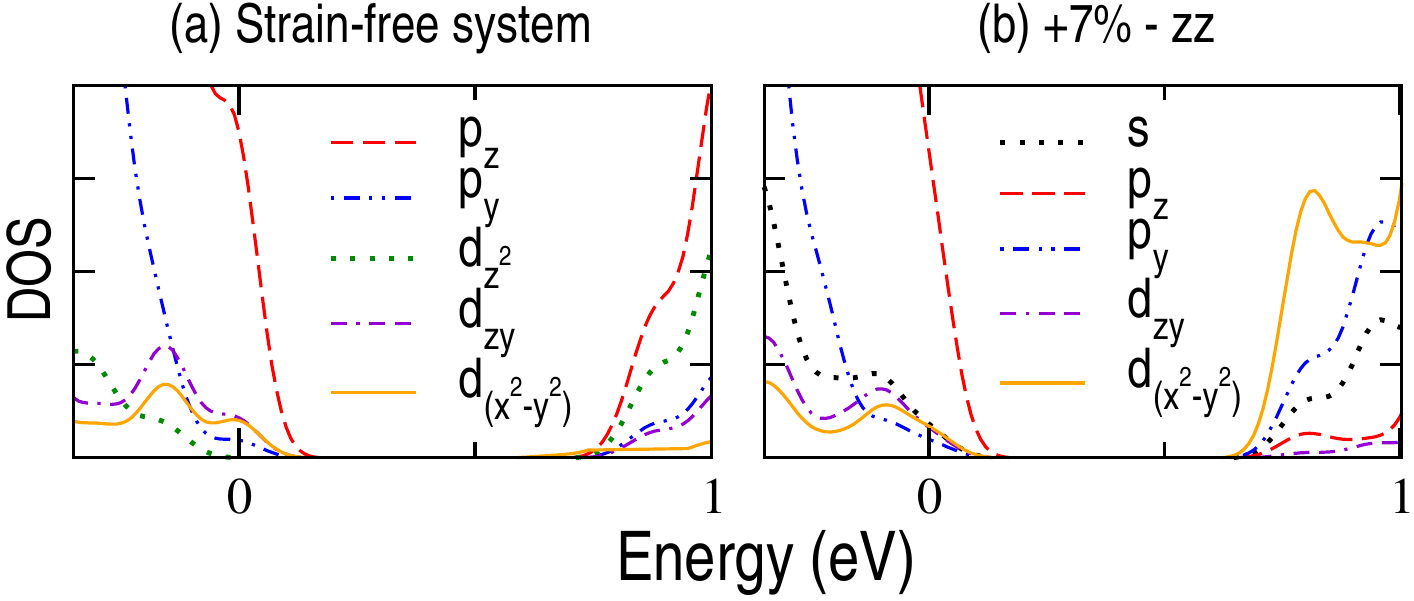}
	\caption{\label{dos} (Color online)  The orbital projection of the near gap electronic density of states of strain-free phosphorene (a), and phosphorene under +7\% tension in $zz$ direction (b).}
\end{figure}

\begin{table*}[bht]
	\caption{\label{table:pdos}
		The projected density of states (PDOS) around the gap area of phosphorene with no strain and when 7\% tensile strain is applied along the $zz$ direction. The percent contribution from each atomic orbital is listed. VBM (CBM) denotes the valence band maximum (the conduction band minimum). The phosphorene $zz$ ($ac$) direction is along $x$ ($y$) axis. }
	\begin{center}
		\begin{tabular}{ c c c c c c c c c c c c }
			\hline
			~System~ & ~~$\vec{k}$~~& ~~states~~&
			~s~ & ~$p_x$~ & ~$p_y$~ & ~$p_z$~ & ~$d_{x^2-y^2}$~ & ~$d_{xy}$~ & ~$d_{zx}$~ & ~$d_{zy}$~ & ~$d_{z^2}$~\\
			\hline
			Strain-free  &  $\Gamma$ & VBM & 2 & 0 & 4  & 73 & 11 & 0 & 0 & 10 & 0  \\
			&           & CBM & 0 & 0 & 15 & 50 & 2  & 0 & 0 & 10 & 23  \\
			\hline
			7\%-zz      &  $\Gamma$ & VBM & 7  & 0 & 6  & 70 & 10 & 0 & 0 & 7 & 0  \\
			&           & CBM & 13 & 0 & 19 & 7  & 55 & 0 & 0 & 2 & 4 \\
			\hline
		\end{tabular}
	\end{center}
\end{table*}

To explain the change of the preferred conducting direction as well as the shift of the gap position in the strained systems, we should note that the stretching of phosphorene under tensile strain along the $zz$ direction is accompanied with its compression in the $ac$ direction, and the enhancement of the inter-planar distance which reflects the negative Poisson ratio of phosphorene along this direction \cite{asgari,strp2}. As a result, the intra- (inter-) planar interactions in the stretched phosphorene system will be slightly improved (weakened).
Figure~\ref{dos} and Table~\ref{table:pdos} represent the projected density of states onto the atomic orbitals for the energy area around the gap at the $\Gamma$  point in the strain-free and +7\% strained phosphorene systems. As the band structure diagrams imply, the effect of strain on the density of states close to the valence band maximum (VBM) is negligible.
On the other hand, the conduction band minimum (CBM) of the strain-free system is dominated by inter-planar (out of plane) $p_z$, $d_{z^2}$, and $d_{zy}$ orbitals, where the contribution of all these inter-planar orbitals is strongly suppressed in the strained system. Instead, by decreasing the energy of the intra-planar orbitals, the CBM of the strained system is dominated by intra-planar $d_{x^2-y^2}$ orbital mixed with $s$, and $p_y$ orbitals. It reveals why the carrier effective mass enhances along the $ac$ direction which has out-of-plane component and decreases along the $zz$ direction which is fully intra-planar.

\begin{figure}
	\centering
	\includegraphics[width=0.75 \linewidth] {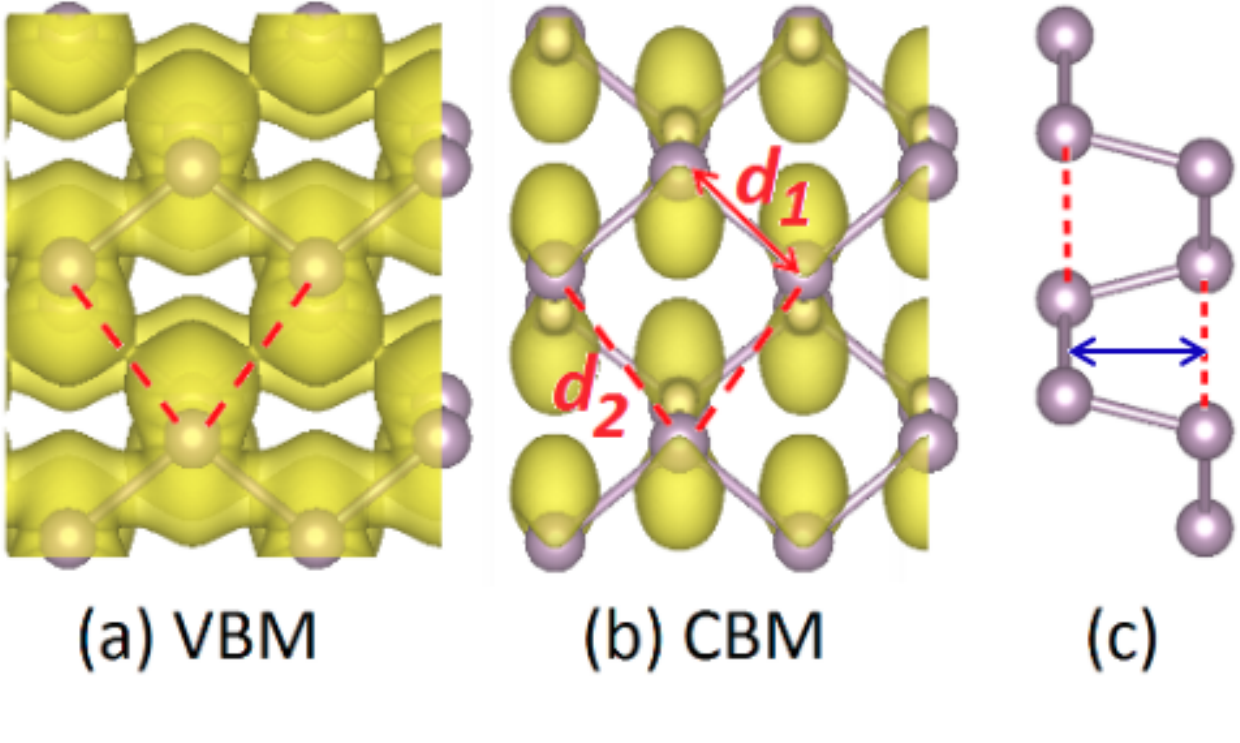}
	\caption{\label{rho} (Color online)  The real-space charge density distribution of the VBM and CBM at the $G$ point (shown in Fig.~\ref{band}(c)). Panel (c) displays a side view of phosphorene in the $ac$ direction. The dashed lines mark bonding between the next-nearest-neighbors (NNN). The blue arrow in panel (c) displays the inter-planar distance in phosphorene. $d_1$ and $d_2$ show the distance between intra-planar NN and NNN atoms.}
\end{figure}

We also perform a charge density analysis based on the Bader's theory to investigate the effect of strain on the charge density distribution in real space, as well as its possible correlation with the electronic band structure properties. This analysis has been successfully used for many crystalline materials \cite{direction}.
The charge density analysis reveals that besides the strong covalent bonding between each phosphorus atom and its three nearest neighbors, the chemical bonds appear between the next-nearest-neighbor (NNN) atoms which are in the same plane. The charge density distribution around the gap point of the phosphorene system under 10\% tensile strain along the $zz$ direction is illustrated in   Fig.~\ref{rho}. This plot indicates that the VBM and CBM at the $G$ point, respectively, originate from the bonding and anti-bonding states between the NNNs.
The tension along the $zz$ direction leads to an increase (strongly decrease) of the distance between the NNs (NNNs) and consequently enhances the bonding between the NNNs. For instance, the bond lengths between the NNs and NNNs in the strain-free phosphorene are respectively $d_1 = 2.2$~\AA~ and $d_2 = 3.3$~\AA, but they change to $d_1 = 2.3$~\AA~ and $d_2 = 3$~\AA under 10\% tension. The bond length reduction between NNNs improves the corresponding exchange energy, based on the Heitler-London Hamiltonian model \cite{strp1,gaas}, and increases (decreases) the energy of the (anti-) bonding states. Accordingly, when the tensile pressure is imposed on phosphorene along the $zz$ direction, the band gap at the $G$ point is gradually reduced and the gap position shifts from $\Gamma$ to $G$ at the tension $\sim$ 9\%.

\begin{figure}
	\centering
	\includegraphics[width=0.33\linewidth]{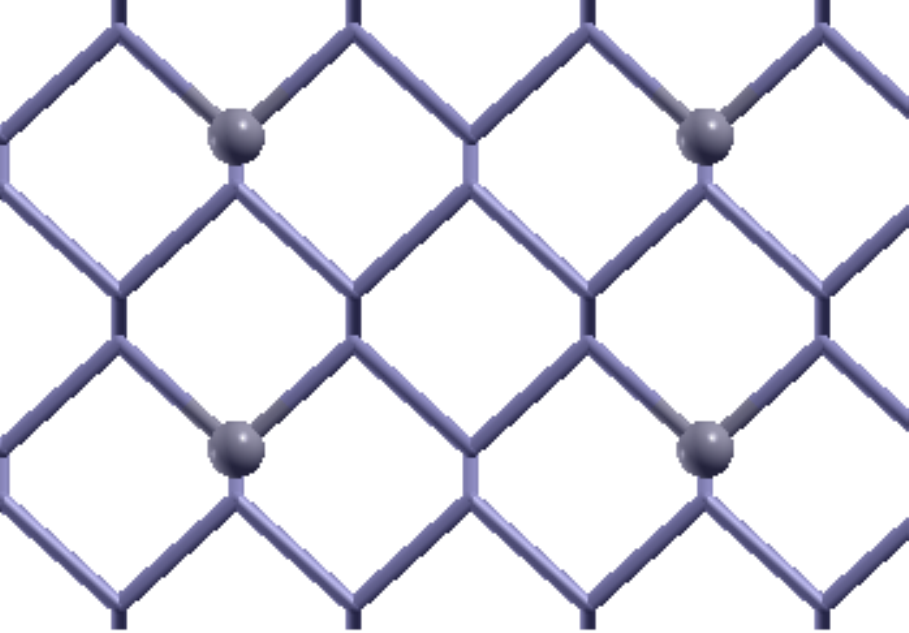}
	\includegraphics[width=1\linewidth]{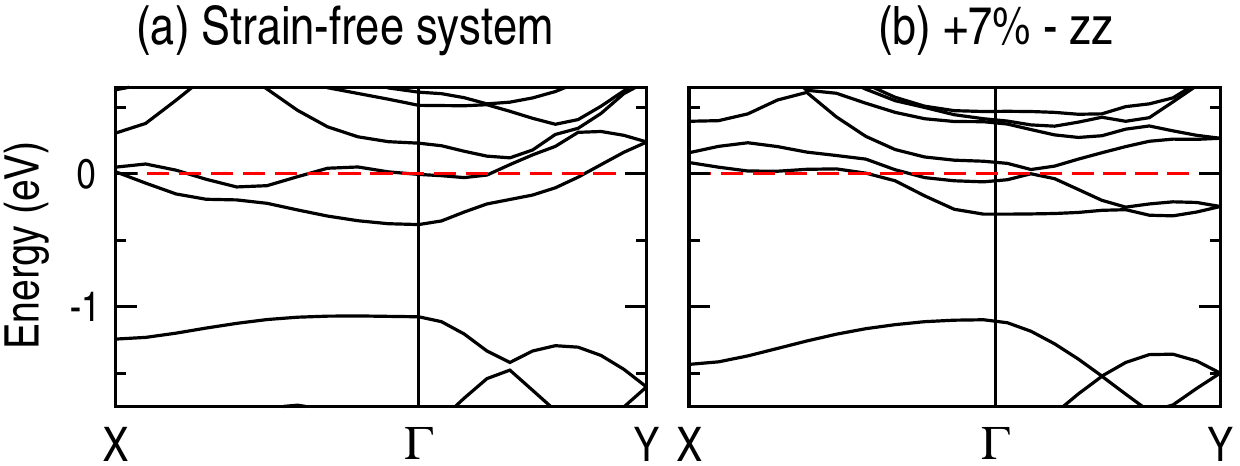}
	\caption{\label{bandS} (Color online)  The electronic band structure diagrams of the S-doped phosphorene system are shown in the top panel of the figure, under free conditions and 7\% tension along $zz$ direction. The dashed lines show the Fermi level, which is set to be zero.}
\end{figure}

The results of the band structure and density of states indicate that the responses of the valence and conduction bands to an applied strain are very different, and the conduction band states are more sensitive than the valence band states against the tensile strain. Furthermore, as the band structure curvature indicates the effective mass of electrons is less than the hole one \cite{strp1}. Therefore, shifting the Fermi surface toward the conduction bands enable us to improve the electromechanical response of the device. This goal is achieved by adding n-type substitutional dopants in the phosphorene system.  Figure~\ref{bandS} shows the configuration of a sulfur (S)-doped phosphorene, as well as the effect of strain on the band structure of this system.
Since the atomic sizes of S and P are close to each other, the lattice distortion of the S-doped system is negligible.
As expected, the band structure around the Fermi level is strongly affected by strain. Same as pristine phosphorene, the band structure dispersion diagrams imply the rotation of the preferred conducting direction from the $ac$ in the strain-free system to the $zz$ in the strained system. This metallic system is suitable for the electrode part of the device.

\subsection{Piezoconductance in phosphorene}\label{sec:tran}

\begin{figure}
	\centering
	\includegraphics[width=1\linewidth] {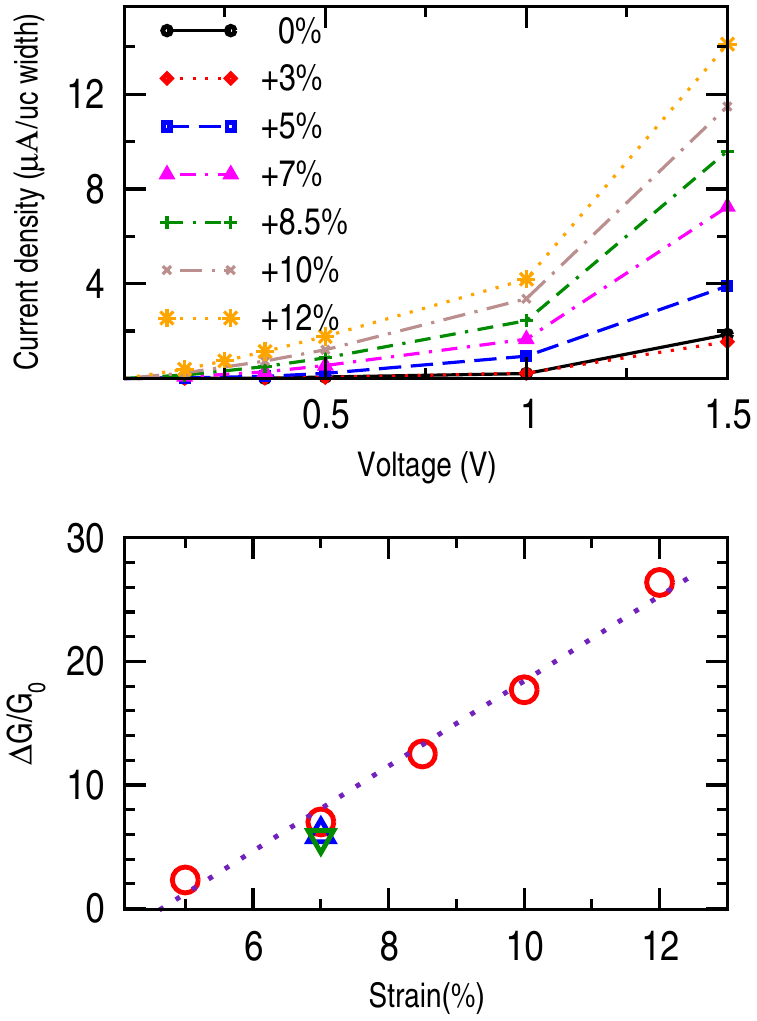}
	\caption{\label{iv-zz}  (Color online) Top panel: $I-V$ characteristics of the proposed phosphorene device (shown in Fig.~\ref{device})  under different tensile strains along $zz$; the applied bias is in $zz$ direction. $uc$ denotes the unit cell.
	Bottom panel: The relative changes of conductance ($G$) versus applied strain over the $zz$ direction. $G_0$ is the conductance of the strain-free system. The triangles show the $\Delta G/G_0$ ratio under the external strain of 7\% for two different devices with 0.75 times shorter and 1.5 times longer scattering regions with respect to that of the reference system ($l_0 \sim 3.5$ nm). }
\end{figure}

In this section, we consider the strain-free phosphorene device as well as different strained systems under an applied bias voltage to the aim of investigating the electromechanical properties of phosphorene. Because of the left-right symmetry of the proposed structures, we just consider the positive biases in the range of $0-1.5$ V. The external tensile strain is applied along the $zz$ direction on the whole device, including the scattering region and electrodes, and it is increased up to 12\%. Having discussed in the earlier section and shown in Fig.~\ref{device}, the S-doped phosphorene is used for the electrode parts which leads to the formation of ohmic electrical barriers at the junctions between electrodes and the central region.

The top panel of Fig.~\ref{iv-zz} displays the $I-V$ characteristics of different strained phosphorene devices as well as a strain-free system under an applied bias along the $zz$ direction.  This figure clearly exhibits the mechanical source of the conductance modulation.  Here similar to the MoS$_2$ device \cite{mos2-piezo}, for a tensile strain larger than $\sim 4-5\%$, the resistance will be reduced by increasing the tension. This is in contrast to graphene, black phosphorous, Si and Ge \cite{gr,blackp2,piezoRrev} piezoresistance devices where the resistance increases under tensile strain. Therefore, it seems that for the former type of materials, piezoconductance is a better label to describe the system.

The electrical resistance of a material depends on its intrinsic properties and its geometry, and it is defined by $R = \frac{l}{\sigma_0 A}$ where $l$ ($A$) is the device length (cross-section area) and $\sigma_0$ is its conductivity. The geometry changes under strain, but in piezoresistance (piezoconductance) materials, the resistance modulation induced by conductivity is several orders of magnitude greater than the geometrical effects.
The conductivity is inversely proportional to the gap size and charge carrier's effective mass. Also, it depends on the density of the charge carriers \cite{mos2-piezo}.
When the tension is applied along the $zz$ direction of phosphorene, the gap size first increases and then continuously decreases (see Fig.~\ref{band}(d)). Moreover, the electron effective mass in this direction is an inverse function of tension \cite{aniso2}.
Note that in the presented device, besides the change of the electronic structure of the central scattering region under strain, the modulation of the electronic structure of the electrodes contributes in the piezoconductance characteristics of the system.
According to the top panel of Fig.~\ref{iv-zz}, for a tensile strain of 3\%, the gap enhancement effect cancels the other effects, and the $I-V$ characteristic of the 3\%
strained system is close to the strain-free one, however, the conductance increases by increasing the tension for the higher values of strain.

The sensitivity of the piezoresistance characteristic is measured by a gauge factor which is determined as the ratio of the relative change of resistance to strain value, GF = $\frac{\Delta R/R_0}{\epsilon}$. Here, similar to the GF defined for a piezoresistance characteristic, we estimate the piezoconductance sensitivity of our device as the ratio of the relative change of conductance to strain value, GF = $\frac{\Delta G/G_0}{\epsilon} = \frac{\Delta I/I_0}{\epsilon}$, where $G$ is the conductance, $G = 1/R$.
It is worth mentioning that the calculated GFs of the usual well-known piezoresistance materials like carbon nanotube, graphene, silicon and germanium \cite{piezoRrev}, as well as the reported GF for MoS$_2$ \cite{mos2-piezo, gf} and black phosphorous systems \cite{blackp2} are obtained for very low strains in the range of $\sim 0.1\%$ where $R \approxeq R_0$, so, we conclude that
$\frac{\Delta R}{R_0} \approxeq \frac{\Delta R}{R}= -\frac{\Delta G}{G_0}$. Therefore, we can compare our GF with the reported GFs for these materials.

To calculate the $\Delta G/G_0$ ratio, which is equal to $\Delta I/I_0$, we use the linear part of the $I-V$ diagrams namely the bias range below 0.5~V. The lower panel of Fig.~\ref{iv-zz} shows the relative change of the conductance (current) as a function of strain.
The value of the GF shows the pressure dependency, and it increases by strain from 46 in 5\% tension to 220 in 12\% tension.
As discussed in Fig.~\ref{band}, the behavior of the gap evolution against strain changes at $\epsilon = +5\%$. Fig.~\ref{iv-zz} indicates that the conductance is enhanced more rapidly and almost linearly versus strain after $\epsilon = +5\%$. In this region, the $\Delta G/G_0$ ratio rises with the slope of $\sim 345$ against the strain.
The reproducibility of the piezoconductance response is investigated by calculating the $\Delta G/G_0$ ratio for two different devices with smaller and larger lengths of the scattering region with respect to that of the reference system ($l_0 \sim 3.5$ nm). The calculations are performed under the applied strain equal to 7\%, and as shown in the lower panel of Fig.~\ref{iv-zz}, the relative change of the conductance is almost independent of the device size.

The predicted gauge factor of our device is one or two orders of magnitude higher than that of suspended graphene membranes \cite{gr},
and it is comparable with the reported value in black phosphorus (122 \cite{blackp2}), MoS$_2$ (148 \cite{mos2-piezo, gf}), and state-of-the-art Si and Ge based piezoresistance devices \cite{piezoRrev}. However, the sensitivity per unit area of phosphorene is higher than that of Si and Ge piezoresistance sensors. Moreover, the holding strain of the current commercial piezoresistance devices is less than 1\%.
Based on the ultrasensitive piezoconductance behavior of phosphorene, its high holding strain limit, and the facilities of the low-dimensional devices, phosphorene has a higher tendency for a wide range of electromechanical applications in high-pressure limits, like in  biological and soft matter deformable systems.
In addition, the enhancement of the conductance and current density, as an experimental observable quantity, in piezoconductance devices has more beneficial effects compared to increasing the resistance of the device which may induce some measurement difficulty due to the current reduction.

\begin{figure}
 	\centering
 	\includegraphics[width=1\linewidth] {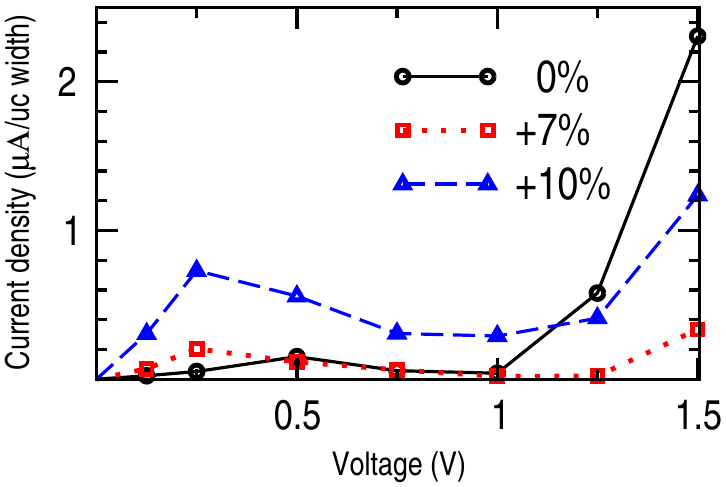}
 	\caption{\label{iv-ac}  (Color online) $I-V$ characteristics of the phosphorene device under tensile strain along the $zz$ direction. The applied bias is in the $ac$ direction. $uc$ denotes the unit cell.}
\end{figure}

Figure~\ref{iv-ac} shows the $I-V$ characteristic of the studied strained systems when the applied bias is perpendicular to the tension and it is along the $ac$ direction. The negative differential resistance (NDR) behavior, where the current decreases through the bias enhancement, is predicted in the system. This phenomenon can be understood by looking at the individual multi-bands along the $ac$  direction of the band structure (see Figs.~\ref{band} and \ref{bandS}). By shifting up and down the discrete and narrow density of states of the left and right electrodes, the bias voltage could provide the conditions that transmission reduces under bias enhancement, leading to the resonant tunneling transport and observed NDR behavior \cite{datta,nour2}.
The peak-to-valley ratio (PVR), which is defined as the ratio of the current at the resonant tunneling peak energy to that at the valley, is obtained around 2.5 - 3.8, which is comparable to the reported values of 7-25 in the $zz$ phosphorene nanoribbons (zPNRs), or $5.5$ in the z-MoS$_2$ NRs  \cite{nour2}, but very small in comparison with the observed PVR value in CdSe quantum dots ($\sim 1000$), or a PVR value of  $50-200$ in defected aGNRs \cite{nour2}. The PVR quantity is a measure of the strength of NDR characteristic.
By increasing the applied strain, due to the strong gap reduction in both $\Gamma$ and $G$ points, the conductance is strongly enhanced at the low biases.
The influence of the carrier effective mass enhancement on the conducting channels of the strained systems is clearly observed in the comparison of the $I-V$ characteristics of this figure with the $I-V$ behavior displayed in the upper panel of Fig.~\ref{iv-zz} where the applied bias is along the $zz$ direction. The $I-V$ characteristics of strain-free systems are comparable with each other, but in the strained systems, the current intensity along the $zz$ direction is stronger than that in the $ac$ direction.

\begin{figure}
  	\centering
  	\includegraphics[width=1.\linewidth] {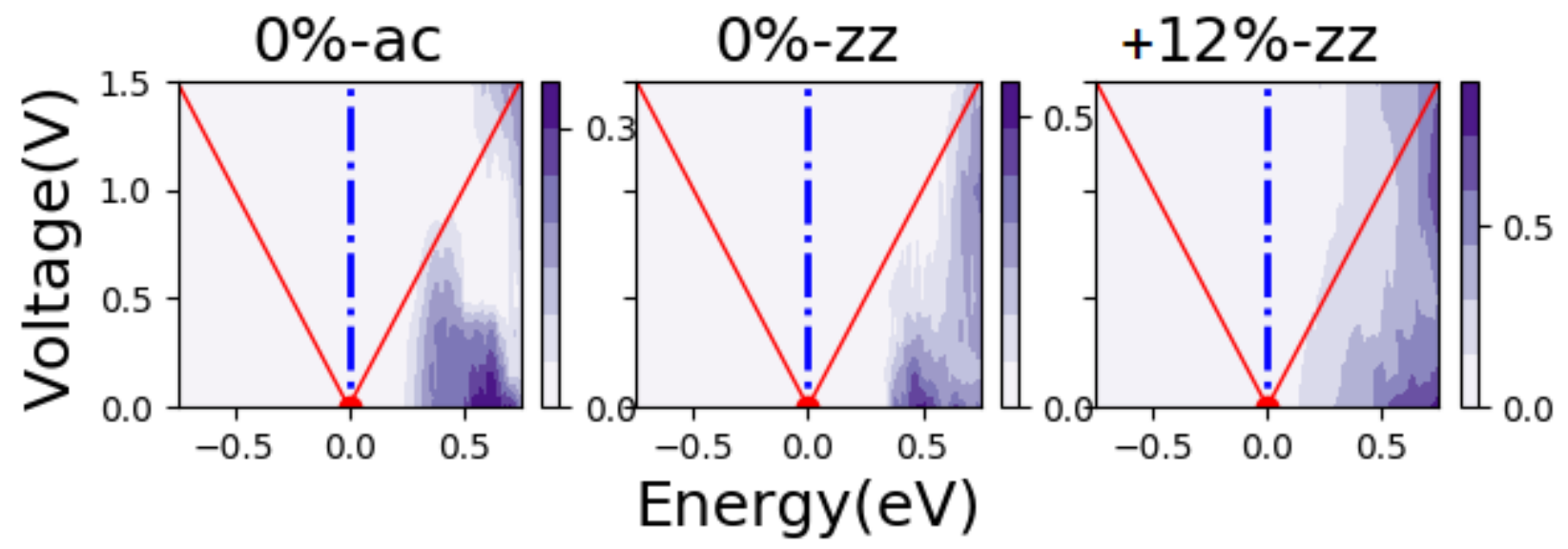}
  	\caption{\label{tran} (Color online) The transmission spectra of strain-free phosphorene system when the applied bias is along the (a) $ac$  and (b) $zz$. (c): Phosphorene under 12\% tensile strain when both tension and applied bias are along the $zz$ direction. The spectra are as functions of charge carrier energy and bias voltage. Two solid red lines determine the bias window in the given voltage. The dashed line shows the Fermi level which is set to be zero.}
\end{figure}

The transmission spectra of the strain-free system under applied biases along both the $ac$ and $zz$ directions, as well as the transmission spectrum of the system under 12\% tensile strain along the $zz$ direction, and applied bias in the $zz$ direction are illustrated in Fig.~\ref{tran}. The spectra are functions of the carrier energy and bias voltage; the solid red line in the negative (positive) energies determines the electrochemical potential of the left (right) electrode, $\mu_{L(R)}$. The energy area between $\mu_l$ and $\mu_R$ is called bias window.
Based on the Landauer formula presented in Eq.~\ref{landaue}, the transmission integration over the bias window is proportional to the current density at the given bias. The transmission in the negative (positive) energies shows the hole (electron) contribution in transport. Therefore, here the dominant part of the conductance comes from electrons. The smaller effective mass of electrons compared to holes in phosphorene \cite{strp1} suggests that electrons have more tendency to participate in the conductivity.

\begin{figure}
  	\centering
  	\includegraphics[width=0.5\linewidth, height=0.9\linewidth, angle= 90] {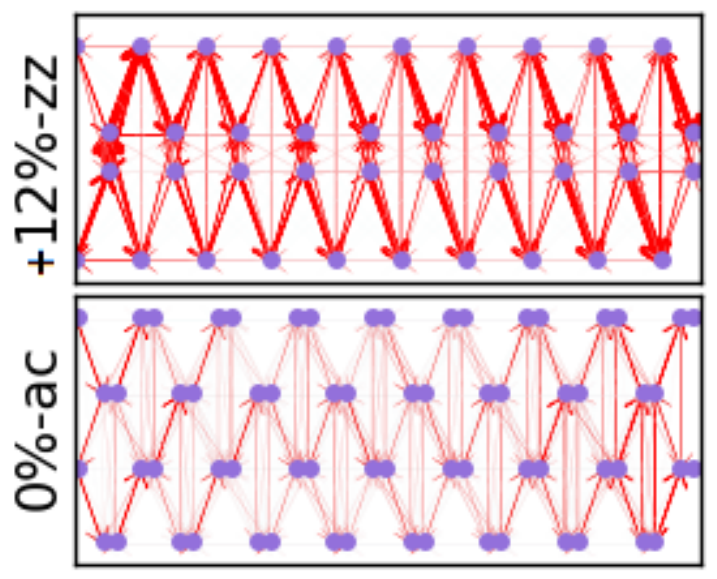}
  	\caption{\label{bndI} (Color online) Visualization of the transmission pathways (or bond current) which show how current flows through the device.
  		The upper panel: both tensile strain (+12\%) and the applied bias (1.5 V) are along $zz$ direction. The lower panel: the strain-free system under an applied bias of 1.5 V along the $ac$  direction. The width of the arrows is related to the magnitude of the local current density. The local currents are totaled over the energy bias window.}
\end{figure}

In order to investigate more details of the charge transport mechanism in our device, we calculate the transmission pathways (also known as bond current) of the strain-free system when the applied bias is along the $ac$ direction, and the transmission pathways of the system under 12\% tension when the electric bias is along the $zz$ direction. The applied bias is equal to $1.5$ V, and the results are displayed in Fig.~\ref{bndI}.

The transmission pathway \cite{handbook, sisl, nour2} is extracted from the splitting of the transmission coefficients into the bond directions and shows how carriers move through the system. The transmission pathway illustrated in Fig.~\ref{bndI} is summed up over the bias window. The directions and width of the arrows are respectively, related to the direction and magnitude of the local current density between the given pair of atoms. Regarding Fig.~\ref{bndI}, the current mainly passes through the chemical bonds, which is called bonding current. In addition, hopping current between further neighboring atoms is observed.
Since the chemical bonds are not along the applied bias, the directions of local currents are not parallel to the applied bias, and as the upper panel of Fig.~\ref{bndI} illustrates, they move through the chemical bonds from the top on the left to the bottom on the right.
This figure is comparable with the calculated transmission pathways in $zz$ phosphorene nanoribbons where the charge carriers pass through the chemical bonds in the vicinity of the $zz$ edges \cite{nour2}.
The conductance of the system displayed in the lower panel of Fig.~\ref{bndI} with the bias along the $ac$ direction, is less than that of the strained system presented in the top panel of the figure; so, the arrows in the lower panel are very narrow.

Finally, we would like to emphasize that the observed piezoconductance behavior arises from the phosphorene band structure properties in electrodes and scattering region. Therefore, changing the dopant concentration of electrodes could be a way to control the current intensity as well as piezoconductance properties of such a phosphorene base device.

\section{Summary}\label{sec:summary}

In this paper, on the basis of $ab~initio$ calculations, we have studied a  2D phosphorene system under tensile strain along the zigzag direction. Due to its puckered structure, phosphorene has remarkable mechanical properties and could sustain tensile strains up to 30\%. In addition,  the phosphorene band structure strongly changes under tension, and the gap size as well as the carrier effective masses are tuned by the applied strain. Based on all these properties, phosphorene has a strong potential for application as a nanoelectromechanical material.

Our results presented in Sec.~\ref{sec:results}, are divided into two parts. First, we have employed charge density analysis techniques based on Bader's \tql Atoms in Molecules\tqr theory to explore the effect of strain in the real space properties of phosphorene, as well as finding the possible connections between the real space and $k$-space behaviors.
Shifting the gap position as well as changing the gap size in the strained phosphorene are understood by forming bonds between the in-plane next-nearest-neighbor phosphorus atoms. Also, the rotation of the preferred conducting direction in strained phosphorene is explained by the enhancement of the intra-planar interactions together with the weakening the inter-planar contribution of the charge density.

Finally, the transport properties of our proposed piezoconductance device have been studied under an applied bias. A strong electromechanical response has been predicted in this system. Our focus of this study is on the relatively high stresses in the range between 5\%-12\%. Based on our results, phosphorene is a promising material for nanoelectromechanical applications in high-pressure limits.
The transmission spectra reveal that transport comes from electrons in the conduction bands. Moreover, the transmission pathways show how electrons propagate through the chemical bonds of the scattering region from the source electrode toward the drain.
The doping concentrations of electrodes are introduced as a tool to control and engineer the piezoconductance properties.

\section*{Acknowledgments}

The numerical simulations are performed using the computational facilities of the School of Nano Science of IPM. Z.N. acknowledges P. Sasanpour for valuable discussions. This work is supported by an Iran Science Elites Federation grant.


\begin{thebibliography}{7}

\bibitem{StrBook}
Y. Sun, S. E. Thompson, and T. Nishida,
\textit{Strain Effect in Semiconductors: Theory and Device Applications}
(Springer, New York, 2010).

\bibitem{nanoSt1}
 A. I. Hochbaum and P. Yang, Chem. Rev.{\bf 110}, 527 (2009).

 \bibitem{nanoSt2}
 X. H. Peng and P. Logan, Appl. Phys. Lett. {\bf 96}, 143119 (2010).

 \bibitem{ph-birth}
 L. Li, Y. Yu, G. J. Ye,	Q. Ge, X. Ou, H. Wu, D. Feng, X. H. Chen and Y. Zhang,
 Nature Nanotech. {\bf9}, 372 (2014).

 \bibitem{aniso1}
 H. Liu, A. T. Neal, Z. Zhu, Z. Luo, X. Xu, D. Tomanek, and P. D. Ye,
 ACS Nano {\bf8}, 4033 (2014).

 \bibitem{ph-app1}
 S. P. Koenig, R. A. Doganov, H. Schmidt, A. H. Castro Neto, and O. Barbaros,
 Appl. Phys. Lett. {\bf104}, 103106 (2014)

 \bibitem{ph-app2}
 M. Buscema, D. J. Groenendijk, S. I. Blanter, G. A. Steele, H. S. J. van der Zant, and A. Castellanos-Gomez,
 Nano Lett., {\bf14}, 3347 (2014).

 \bibitem{ph-app3}
 F. Xia, H. Wang	and Y. Jia,
 Nature Commun. {\bf5}, 4458 (2014).

\bibitem{aniso2}
R. Fei and L. Yang,
Nano Lett. {\bf14}, 2884 (2014).

 \bibitem{PExcExp}
 X. Wang, A. M. Jones, K. L. Seyler,	V. Tran, Y. Jia, H. Zhao, H. Wang, L. Yang, X. Xu and F. Xia,
 Nature Nanotech. {\bf 10}, 517 (2015)

\bibitem{nour1}
Z. Nourbakhsh and R. Asgari,
Phys. Rev. B {\bf94}, 035437 (2016).

\bibitem{strp1}
X. Peng, Q. Wei, and A. Copple,  Phys. Rev. B {\bf90}, 085402 (2014).
Q. Wei and X. Peng,  App. Phys. Lett. {\bf 104}, 251915 (2014).

\bibitem{gr}
A. D. Smith, F. Niklaus, A. Paussa, S. Schroder, A. C. Fischer, M. Sterner, S. Wagner, S. Vaziri, F. Forsberg, D. Esseni, M. Ostling, and M. C. Lemme,
ACS Nano {\bf 10(11)}, 9879 (2016).
A. D. Smith, F. Niklaus, A. Paussa, S. Vaziri, A. C. Fischer, M. Sterner, F. Forsberg, A. Delin, D. Esseni, P. Palestri, M. Östling, and M. C. Lemme,
Nano Lett. {\bf 13(7)}, 3237 (2013).

\bibitem{mos2}
S. Bertolazzi, J. Brivio, A. Kis,
ACS Nano, {\bf 5(12)}, 9703 (2011).

\bibitem{asgari}
M. Elahi, K. Khaliji, S.M. Tabatabaei, M. Pourfath, and R. Asgari, Phys. Rev. B {\bf 91}, 115412 (2015).

\bibitem{strp2}
A. S. Rodin, A. Carvalho, and A. H. Castro Neto, Phys. Rev. Lett. {\bf 112}, 176801 (2014).

\bibitem{strp3}
Y. Li, Sh. Yang, and J. Li,
J. Phys. Chem. C  {\bf 118 (41)}, 23970 (2014).

\bibitem{strp4}
L. Seixas, A. S. Rodin, A. Carvalho, and A. H. Castro Neto,
Phys. Rev. B {\bf 91}, 115437 (2015).

\bibitem{blackp2}
Z. Zhang, L. Li, J. Horng, N. Zhou Wang, F. Yang, Y. Yu, Y. Zhang, G. Chen, K. Watanabe, T. Taniguchi, X. Hui Chen, F. Wang, and Y. Zhang,
Nano Lett.  {\bf 17(10)}, 6097 (2017).

\bibitem{negPhonon}
B. Sa, Y.-L. Li, J. Qi, R. Ahuja, and Zh. Sun,
J. Phys. Chem. C {\bf 118 (46)}, 26560 (2014).

\bibitem{piezoRrev}
A. A. Barlian, W.-T. Park, J. R. Mallon, A. J. Rastegar, and B. L. Pruitt,
Proc IEEE {\bf 97(3)}, 513 (2009). 

\bibitem{smith}
C. S. Smith, Phys Rev {\bf 94}, 42 (1954).

\bibitem{mos2-piezo}
S. Manzeli, A. Allain, A. Ghadimi, and A. Kis,
Nano. Lett. {\bf15(8)}, 5330 (2015).

\bibitem{blackp1}
D. M. Warschauer,
J. App. Phys. {\bf 35}, 3516 (1964).

\bibitem{k-sh}
W. Kohn and L. J. Sham,
Phy. Rev. {\bf 140}, A1133 (1965).

\bibitem{negf}
N. E. Dahlen, A. Stan, and R. van Leeuwen,
J. Phys.: Conference Series {\bf 35}, 324 (2006).

\bibitem{landauer}
R. Landauer, Philos. Mag. {\bf21}, 863 (1970).

\bibitem{exp1}
T. Frederiksen, G. Foti, F. Scheurer, V. Speisser, and G. Schull, Nat. Commun. {\bf5}, 3659 (2014).
G. Schull, T. Frederiksen, M. Brandbyge, and R. Berndt,
Phys. Rev. Lett. {\bf103}, 206803 (2009).
M. L. N. Palsgaard, N. P. Andersen, and M. Brandbyge,
Phys. Rev. B {\bf91}, 121403(R) (2015).
J. Lagoute, F. Joucken, V. Repain, Y. Tison, C. Chacon, A. Bellec, Y. Girard, R. Sporken, E. H. Conrad, F. Ducastelle, M. Palsgaard, N. P. Andersen, M. Brandbyge, and S. Rousset,
Phys. Rev. B {\bf91}, 125442 (2015).
N. L. Schneider, N. Neel, N. P. Andersen, J. T. Lu, M. Brandbyge, J. Kroger, and R. Berndt,
J. Phys.: Cond. Mat {\bf 27}, 015001 (2015).

\bibitem{bader}
R. Bader, \textit{Atoms in Molecules: A Quantum Theory}
(Oxford University Press, UK, 1994).

\bibitem{transiesta}
M. Brandbyge, J.-L. Mozos, P. Ordejon, J. Taylor, and K. Stokbro,
Phys. Rev. B {\bf65}, 165401 (2002).

\bibitem{handbook}
J. E. Morris, K. Iniewski,
\textit{Nanoelectronic Device Applications Handbook}
(CRC Press, Cleveland, OH, 2013).

\bibitem{siesta}
J. M. Soler, E. Artacho, J. D. Gale, A. Garcia, J. Junquera, P. Ordejon, D. Sanchez-Portal,
J. Phys.: Cond. Mat {\bf14}, 2745 (2002).
http://departments.icmab.es/leem/siesta/

\bibitem{pbe}
J. P. Perdew, K. Burke, and M. Ernzerhof,
Phys. Rev. Lett. {\bf 77}, 3865 (1996).

\bibitem{m-p}
H. J. Monkhorst and J. D. Pack,
Phys. Rev. B. {\bf13}, 5188 (1976).

\bibitem{aimall}
http://aim.tkgristmill.com

\bibitem{tbtran}
N. Papior, N. Lorente, T. Frederiksen, A. Garcia, and M. Brandbyge,
Comput. Phys. Commun. {\bf 212}, 8 (2016).

\bibitem{sisl}
N. R. Papior,
https://doi.org/10.5281/zenodo.597181;
https://github.com/zerothi/sisl.
More information about the bond and atomic current and their mathematical equations could be find in sisl documentation.
http://zerothi.github.io/sisl/docs/latest/api-generated/sisl.io.tbtrans.phtavncSileTBtrans.html

\bibitem{direction}
Z. Nourbakhsh, S. J. Hashemifar, H. Akbarzadeh,
J. Alloy. Compd. {\bf579}, 360 (2013).
T. E. Jones, M. E. Eberhart, and D. P. Clougherty,
Phys. Rev. Lett. {\bf100}, 017208 (2008).

\bibitem{gaas}
X. Peng and A. Copple,
Phys. Rev. B {\bf87}, 115308 (2013).

\bibitem{datta}
S. Datta,
\textit{Quantum Transport: Atom to transistor}
(Cambridge University Press, London, 2005).

\bibitem{gf}
In Ref. \cite{mos2-piezo}, Manzeli $et~al.$ used the relative change of resistance versus strain to show the piezoresistance sensitivity of MoS$_2$. They studied the transport behavior of the system for  tensile strains up to 7\%. However, their reported GF could be correct just for the strain range less than $0.2\%$ as they considered the resistance as a function of strain by the formula $R = R_0 exp(\alpha \ep) \approxeq  R_0(1+\alpha \ep)$ where $\alpha = -148$ and the expansion is  acceptable just for strains less than $0.2\%$.

\bibitem{nour2}
Z. Nourbakhsh and R. Asgari,
Phys. Rev. B {\bf97}, 235406 (2018).


\end{thebibliography}
\end {document}